\documentclass[aps,pre,preprint,groupedaddress,superscriptaddress]{revtex4-1}%
\usepackage{amsfonts}
\usepackage{amsmath}
\usepackage{amssymb}
\usepackage{graphicx}
\usepackage{natbib}%
\usepackage{caption}
\usepackage[labelsep=quad,indention=10pt]{subfig}
\providecommand{\U}[1]{\protect\rule{.1in}{.1in}}
\begin{document}
\preprint{ }
\title{Effective potentials in polyelectrolyte solutions}

\author{Sandipan Dutta}
\affiliation{Asia Pacific Center for Theoretical Physics, Pohang, Gyeongbuk, 790-784, Korea}
\author{and Y.S. Jho}
\email{ysjho@apctp.org}
\affiliation{Asia Pacific Center for Theoretical Physics, Pohang, Gyeongbuk, 790-784, Korea}
\affiliation{Department of Physics, Pohang University of Science and Technology, 790-784, Korea}

\begin{abstract}
Using Poisson-Boltzmann equation and linear response theory, we derive an effective interaction potential 
due to a fixed charge distribution in a solution containing polyelectrolytes and point salt. We obtain an
expression for the effective potential in terms of static structure factor using the integral 
equation theories. To demonstrate the theory we apply it to
Gaussian and rod-like polyelectrolytes and make connections to earlier theoretical works in some exact limits.
We explore the role of both intra and inter polymer correlations, and the geometry of the polymers
in the development of attractive regions in the effective potential as well as their effects on the 
screening lengths.
\end{abstract}

 \pacs{}
%\keywords{}

\maketitle

\section{Introduction}

Polyelectrolyte solutions are of great interest both in the field of science as well as industry. The electrostatic interactions
of charged biopolymers with proteins and membranes are the underlying mechanisms of many biological processes \cite{C2SM27002A,
cooper2005polyelectrolyte,katchalsky1964polyelectrolytes}. In the industry
they have wide range of applications from gels and surfactants to waste water treatment. Polyelectrolyte mediated interactions 
have been well-studied in the field of colloidal science also \cite{van2013polymer,becker2012proteins,turgeon2007protein,
cooper2005polyelectrolyte,krakoviack2004integral}. Many experiments have been performed to study the different phases 
of colloid-polymer mixtures \cite{turgeon2007protein,dickinson2006colloid,sennato2012aggregation,chollakup2013polyelectrolyte,
pelaez2010structure,spruijt2010binodal}. Recently new kind of phenomena like overcharging, charge reversal and 
like-charge attractions in the presence of polyelectrolytes have attracted renewed attention to the study of polyelectrolyte mediated 
interactions \cite{hansen2000effective,rouzina1996macroion,perel1999screening,ha1999counterion,podgornik1998charge,dutta2015adsorption,
dutta2015strong}. Polyelectrolytes are also known to cause charge bridging and depletion interactions among colloidal particles 
\cite{netz2003neutral,dobrynin2008theory,dobrynin2005theory,van1988polyelectrolytes,aakesson1989electric,miklavic1990interaction,
podgornik1991electrostatic,podgornik1993variational,woodward1994forces,huang2004bridging}.

Many theoretical studies have been made on the polyelectrolyte mediated interactions, and have found a rich phase diagram depending
on the nature of the solution, pH level and salt concentrations \cite{messina2004polyelectrolyte,ryden2005monte,
jonsson2001polyelectrolyte,carrillo2007molecular,dias2005polyion,carlsson2001monte,blaak2012complexation,ulrich2005complexation,
turesson2007interactions,ulrich2011formation,truzzolillo2010interaction,antypov2005incorporation,PhysRevLett.111.168303}.
Khokhlov and Khachaturian \cite{khokhlov1982theory} obtained a diagram 
of states for weakly charged polyelectrolytes depending their concentration using an effective potential (EP) approach. Borue and Erukhimovich
\cite{boryu1988statistical} obtained a screened Coulomb potential of a test charge in weakly charged polyelectrolyte solution within the random
phase approximation and found that the potential has an oscillatory regime specially in poor solvents. By employing a 
variational field theoretic calculation to treat monomer density correlations, Muthukumar \cite{muthukumar1996double} derived an EP
between the segments of the polymers. The EP has a short ranged attraction due to the entropy and the connectivity of the chains
and a longer ranged repulsion from the electrostatic interactions. Borukov, Andelman and Orland \cite{borukhov1999effect} described the effects of 
adsorbed polymers on the inter-colloidal forces using a self-consistent field theory. They however predicted a short ranged repulsion
and long ranged attractions between the colloid surfaces. Simulations by Turesson \textit{et al.} \cite{turesson2007interactions} however show that many different 
kind of interactions like short-ranged attractions, longer ranged repulsions and at larger distances weak oscillatory decaying 
interactions are possible in polyelectrolyte solutions. Pryamitsyn and Ganesan \cite{pryamitsyn2014interplay} have 
recently modeled the short ranged attractions in the EP in polyelectrolyte-nanoparticle systems by depletion forces 
and the long range repulsions by the Debye-Huckel potentials. Most of these theories are restricted mainly to the mean field (weak coupling)
regime. Our objective is to develop a formalism that would enable us to investigate the role of correlations in the effective 
interactions mediated by the polyelectrolytes.

In this work we study the effective interactions due to a fixed charge distribution in a polyelectrolyte solution 
using the integral equation theory developed by the current authors \cite{dutta2015shell}. Using the linear response theory together with the Poisson-
Boltzmann equation, we derive an expression for the effective interactions due to the fixed charge distribution 
in the polyelectrolyte solution. In the course
of derivation we obtain a static form of the fluctuation-dissipation theorem \cite{hansen1990theory} for polymers which was known in 
some other form \cite{hayter1980correlations}.
In the mean field approximation, our EP still has a contribution from the configurational entropy of a single polymer
which gets stronger when the polymer length or charge increases. Because of this contribution the EP develops a
short ranged attraction as previously found by Muthukumar \cite{muthukumar1996double}. When the inter-polyelectrolyte correlations,
calculated using the Laria, Wu, and Chandler (LWC) theory \cite{laria1991reference,shew1997integral}, is taken into account the attractions in the EP 
become even stronger and for larger polymers or stronger Coulomb couplings the EP develop an oscillatory behavior similar to the ones obtained in 
the simulations \cite{turesson2007interactions}. We illustrate our model for the case of rod-like and Gaussian polyelectrolytes in presence of a point
test charge. In particular we explore the origin of the attractive regions in the EP and the screening of the test charge potential due to 
these polyelectrolytes.

The paper is organized as follows. In Section \ref{Sec1} we derive an EP due to a fixed charge distribution starting from the Poisson-Boltzmann equation
along with the integral equation theories \cite{dutta2015shell,shew1997integral,schweizer1994prism}.
We look at the mean field limit of the EP and study its dependence 
on the geometry of the polymers and the strength of the Coulomb interactions in Section \ref{Sec2}. 
In Section \ref{Sec3} we calculate the polymer-polymer correlations self-consistently 
using the PRISM and LWC equations and use them to obtain the EP beyond the mean field approximation.
In both the mean field and the correlated cases we investigate the screening by the polyelectrolytes though the Debye length and study its
dependence on the polymer length, monomer length and the polymer charge. We discuss the regime of validity of our model and its
possible generalizations in Section \ref{Sec4}.

\section{The Effective potential}
\label{Sec1}

\begin{figure*}[h]
        \centering
           \subfloat{%
              \includegraphics[trim={7cm 4cm 0 2cm},clip,height=6.2cm]{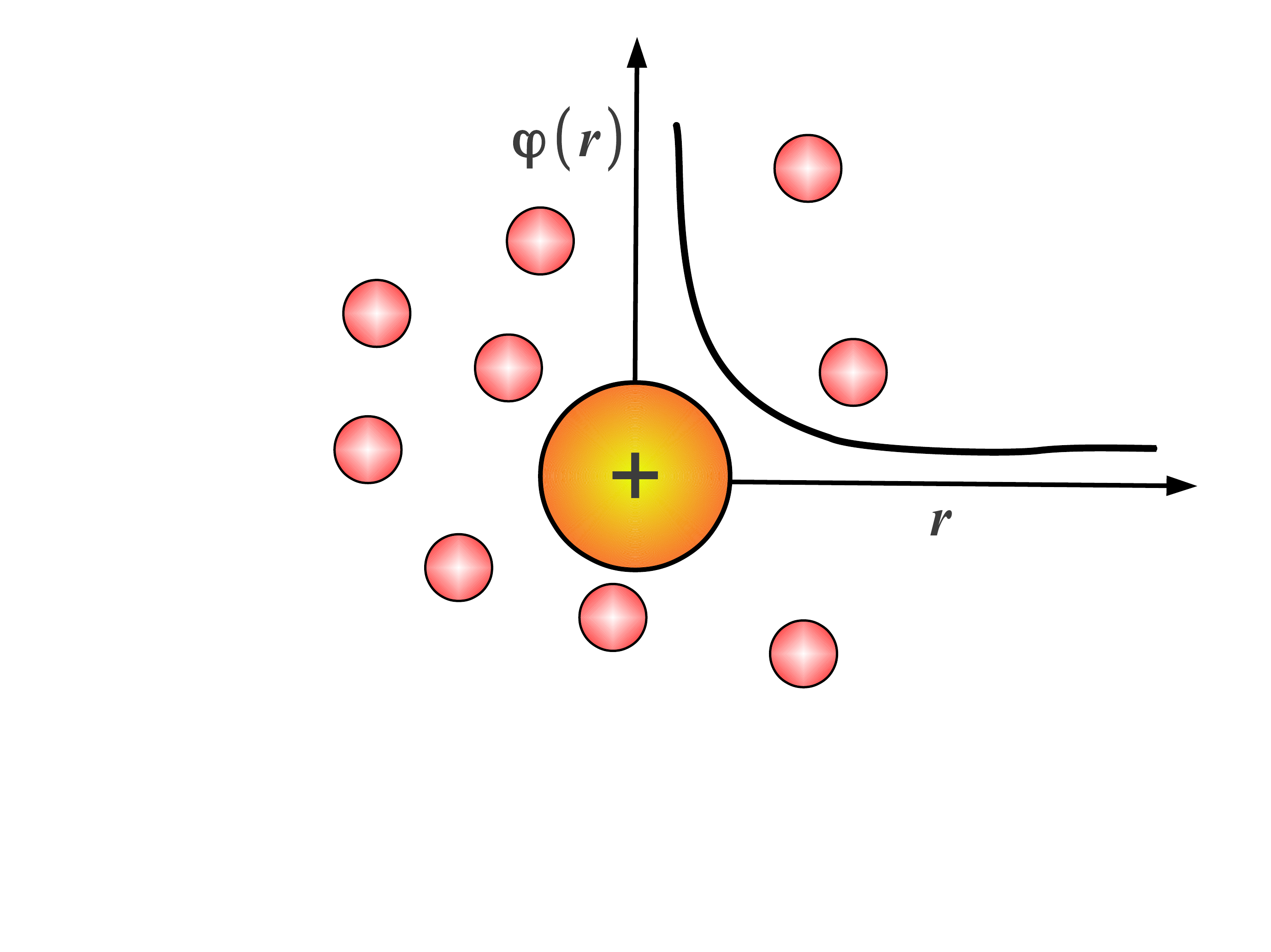}%
           }
           \subfloat{%
              \includegraphics[trim={6cm 2.5cm 0 2cm},clip,height=6.2cm]{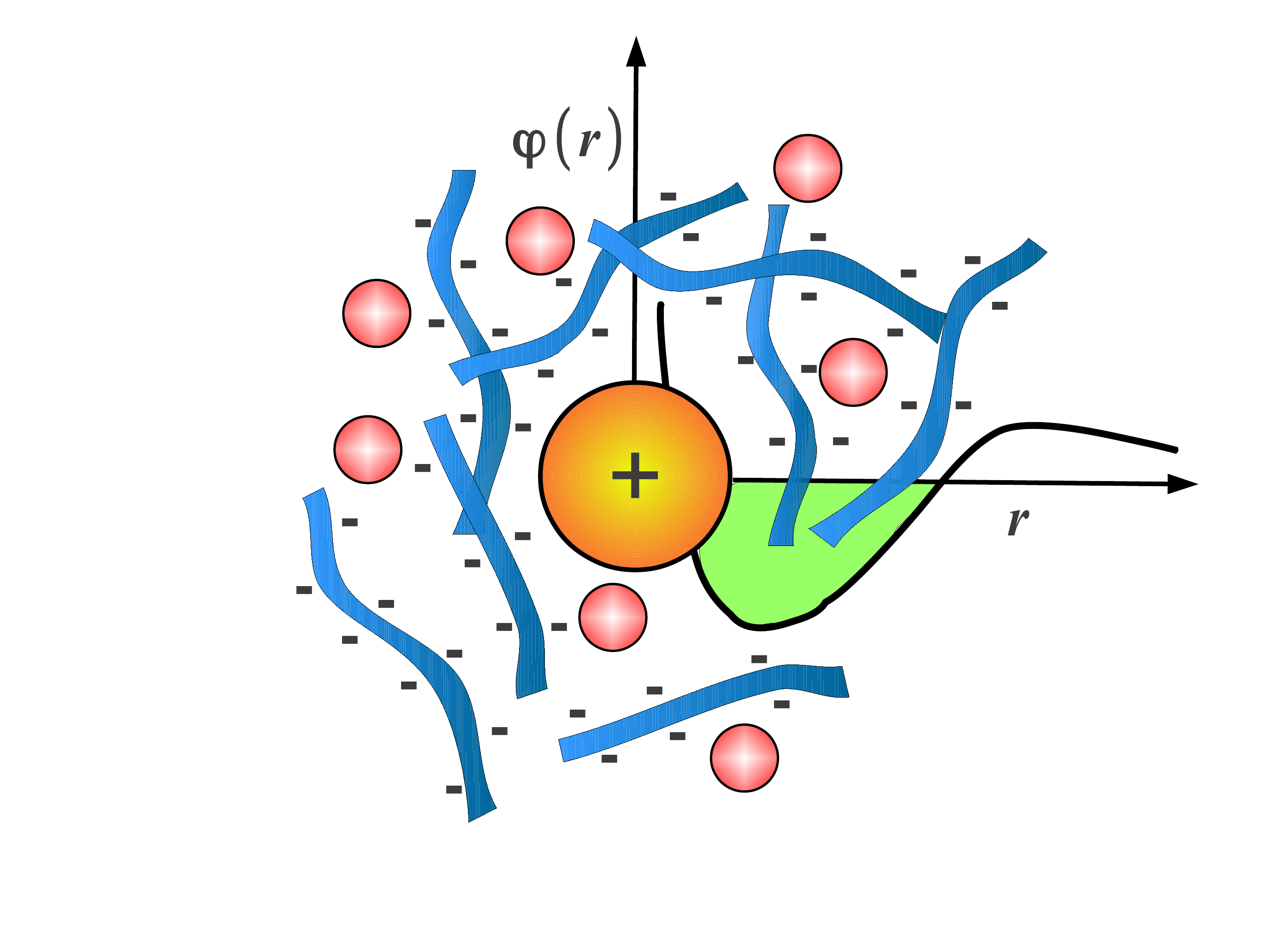}%
           }
           \caption{Schematic diagram of the effective potential due to a test charge in presence of (a) point salt ions and 
           (b) polyelectrolytes and point salt ions. } 
           \label{Fig0}
 \end{figure*}

Consider a system of $N_p$ polyelectrolytes, each consisting of $L$ monomers of length $\sigma$ and charge $q_p$ uniformly smeared over each monomer.
The diameter of the monomers is assumed to be the same as the monomer length.
Other $N_s$ point salt ions of charge $q_s$ are also present in the system along with a fixed charge distribution $\rho_{f}(\mathbf{r})$. 
The densities of the point ions and monomers are given by 
$n_s = N_s/\Omega$ and $n_p = N_pL/\Omega$, where $\Omega$ is the volume of the system. The electrostatic Coulomb interactions are denoted by  
$v(\vert\mathbf{r}-\mathbf{r}^{\prime}\vert) = 1/\epsilon\vert\mathbf{r}-\mathbf{r}^{\prime}\vert$, where $\epsilon$ is the dielectic constant 
of the solvent. The schematic diagram of the system is shown in Figure \ref{Fig0}-(b). The Hamiltonian of the system is
\begin{align}
 H_N = \sum_iH_i^0 + \frac{1}{2}\int d\mathbf{r}\int d\mathbf{r}^{\prime}\left[q_p\hat{\rho}_p(\mathbf{r})+\rho_f(\mathbf{r})+q_s\hat{\rho}_s(\mathbf{r})\right]
 v(\vert\mathbf{r}-\mathbf{r}^{\prime}\vert)\left[q_p\hat{\rho}_p(\mathbf{r}^{\prime})+\rho_f(\mathbf{r}^{\prime})+
 q_s\hat{\rho}_s(\mathbf{r}^{\prime})\right],
 \label{eq1.1}
\end{align}
where $\hat{\rho}_p(\mathbf{r})$, $\hat{\rho}_f(\mathbf{r})$ and $\hat{\rho}_s(\mathbf{r})$ are the densities of the polyelectrolytes,
fixed charge and the salt ions respectively. $H_0^i$ is the single polymer Hamiltonian which contains the information about the bending rigidity,
connectivity and structure of the polymer. The densities of the monomers and the salt are given by
\begin{align}
 \hat{\rho}_p(\mathbf{r}) & = \sum_{i=1}^N\int_0^L ds\delta(\mathbf{x}_i(s)-\mathbf{r}),  \label{eq1.2}\\
 \hat{\rho}_s(\mathbf{r}) & = \sum_{i=1}^{N_s}\delta(\mathbf{r}_i-\mathbf{r}),
 \label{eq1.3}
\end{align}
respectively. 

Since we are mostly interested in the role of the polymers, we convert everything to dimensionless quantities by scaling with respect to the
polymer quantities. We scale the distances with respect to the average distances between the monomers $r_0$: $\widetilde{\mathbf{r}} = \mathbf{r}/r_0$, 
where $r_0$ is defined in terms of the monomer density $n_p$ by $4\pi n_p r_0^3/3 = 1$. Similarly the momentum 
vectors are scaled like $\widetilde{\mathbf{k}} = \mathbf{k}r_0$. The densities are made dimensionless by $\widetilde{\hat{\rho}}(\widetilde{\mathbf{r}})
= \hat{\rho}(\mathbf{r})r_0^3$. Similarly the dimensionless potential is $\Gamma\widetilde{v}(\vert\widetilde{\mathbf{r}}-\widetilde{\mathbf{r}}^{\prime}\vert)
= \frac{\beta q_p^2}{\epsilon r_0}v(\vert\mathbf{r}-\mathbf{r}^{\prime}\vert)$, with $\Gamma = \frac{\beta q_p^2}{\epsilon r_0}$. The charges are scaled by the 
monomer charge of the polymer $\widetilde{q} = q/q_p$. Hence for the polymers the scaled charge is $\widetilde{q}_p = 1$ and
the density $\widetilde{n}_p = 3/4\pi$. The dimensionless Hamiltonian reads
\begin{align}
 \widetilde{H}_N = \sum_i\widetilde{H}_i^0 + \frac{1}{2}\int d\widetilde{\mathbf{r}}\int d\widetilde{\mathbf{r}}^{\prime}
 \left[\widetilde{\hat{\rho}}_p(\widetilde{\mathbf{r}})+\widetilde{\rho}_f(\widetilde{\mathbf{r}})+\widetilde{q}_s\widetilde{\hat{\rho}}_s(\widetilde{\mathbf{r}})\right]
 \Gamma v(\vert\widetilde{\mathbf{r}}-\widetilde{\mathbf{r}}^{\prime}\vert)\left[\widetilde{\hat{\rho}}_p(\widetilde{\mathbf{r}}^{\prime})+\widetilde{\rho}(\widetilde{\mathbf{r}}
 ^{\prime})+ \widetilde{q}_s\widetilde{\hat{\rho}}_s(\widetilde{\mathbf{r}}^{\prime})\right].
 \label{eq1.4}
\end{align}
In the rest of the discussions we only use the dimensionless quantities and drop the $\widetilde{...}$ in their notations.
The canonical partition function with the above Hamiltonian given is
\begin{equation}
 Z = \int \prod\limits_{i=1}^{N_s}d\mathbf{r}_i\prod\limits_{j=1}^{N}\mathcal{D}\mathbf{x}_j\exp\left(-H_N
 [\{\mathbf{r}_i\},\{\mathbf{x}_j\}]\right).
 \label{eq1.5}
\end{equation}
Performing the Hubbard-Stratonovich transformation \cite{stratonovich1957method,hubbard1959calculation} we introduce a field $\phi$ and the partition 
function transforms to \cite{fredrickson2013equilibrium}
\begin{equation}
 Z[\phi] = \int \mathcal{D}\phi \exp\left(-\mathcal{H}[\phi]\right),
 \label{eq1.6}
\end{equation}
where 
\begin{equation}
 \mathcal{H}[\phi] = \frac{1}{2\Gamma}\int d\mathbf{r}\int d\mathbf{r}^{\prime}\phi(\mathbf{r}) v^{-1}\vert\mathbf{r}-\mathbf{r}^{\prime}\vert)
 \phi(\mathbf{r}^{\prime}) + N_s\ln\mathfrak{z}_s[q_s\phi] + N_p\ln\mathfrak{z}_p[\phi] + \rho_f(\mathbf{r})\phi(\mathbf{r}).
 \label{eq1.7}
\end{equation}
In the mean field approximation using $\delta\mathcal{H}[\phi]/\delta\phi(\mathbf{r}) = 0$ we obtain the non-linear Poisson-Boltzmann equation
\begin{equation}
 \frac{1}{\Gamma}\nabla^2\phi(\mathbf{r}) = \rho_p(\mathbf{r};\phi) + \rho_s(\mathbf{r};q_s\phi) + \rho_f(\mathbf{r}).
 \label{eq1.8}
\end{equation}
$\rho_p(\mathbf{r};\phi) = -N_p\delta\mathfrak{z}_p[\phi]/\delta\phi(\mathbf{r})$ and $\rho_s(\mathbf{r};\phi)= -N_s\delta\mathfrak{z}_s[\phi]
/\delta\phi(\mathbf{r})$ are the polymer and salt densities respectively in the external potential $\phi$.
We write the polymer and salt densities in the following form
\begin{align}
 \rho_s(\mathbf{r};\phi) & = \int d\mathbf{r}^{\prime} \chi_s(\vert\mathbf{r},\mathbf{r}^{\prime}\vert)\phi(\mathbf{r}^{\prime}),\label{eq1.9}\\
 \rho_p(\mathbf{r};\phi) & = \int d\mathbf{r}^{\prime} \chi_p(\vert\mathbf{r},\mathbf{r}^{\prime}\vert)\phi(\mathbf{r}^{\prime}),
 \label{eq1.10}
\end{align}
where $\chi_s(\vert\mathbf{r},\mathbf{r}^{\prime}\vert)$ and $\chi_p(\vert\mathbf{r},\mathbf{r}^{\prime}\vert)$ are the response functions 
of the salt and polymers respectively \cite{hansen1990theory}.
For simplicity we use the response functions for uniform systems. Taking the Fourier transform of equation \eqref{eq1.8}
we solve for the potential $\phi(\mathbf{r})$ in presence of the fixed charge distribution by
\begin{equation}
 \hat{\phi}(\mathbf{k}) = \frac{\Gamma\hat{\rho_f}(\mathbf{k})}{k^2 - 4\pi\Gamma q_s\hat{\chi}_s(\mathbf{k}) - 4\pi\Gamma \hat{\chi}_p(\mathbf{k})}.
 \label{eq1.11}
\end{equation}
$\phi(r)$ is the effective potential due to the fixed charge distribution in the presence of the point salt and the polyelectrolytes.

The response function of the polymers, $\chi_p(\vert\mathbf{r}-\mathbf{r}^{\prime}\vert)$ is obtained by perturbing the system by a small external 
potential $\delta\phi(r)$ and the corresponding change in the density of the system \cite{ichimaru1982strongly} 
\begin{equation}
 \delta\hat{\rho}_p(\mathbf{k}) = \hat{\chi}_p(\mathbf{k})\delta\hat{\phi}(\mathbf{k}).
 \label{eq1.12}
\end{equation}
We use the density equation obtained by the current authors based on the reference interaction site model (RISM) theory of Chandler
\cite{chandler1986density2,chandler1986density1} which relates the density to the external potential by
\begin{align}
 \ln\left(\rho_p(\mathbf{r})\lambda^3/z\right)  = -\int d\mathbf{r}^{\prime}d\mathbf{r}^{\prime\prime}\omega(\vert\mathbf{r}-\mathbf{r}^{\prime}\vert)
 \phi(\vert\mathbf{r}^{\prime}-\mathbf{r}^{\prime\prime}\vert)\omega(r^{\prime\prime}) + \int d\mathbf{r}^{\prime}d\mathbf{r}^{\prime\prime}
 d\mathbf{r}^{\prime\prime\prime}&\omega(\vert\mathbf{r}-\mathbf{r}^{\prime}\vert)c(\vert\mathbf{r}^{\prime}-\mathbf{r}^{\prime\prime}\vert)\times
 \nonumber\\& \rho_p(\vert\mathbf{r}^{\prime\prime}-\mathbf{r}^{\prime\prime\prime}\vert),
 \label{eq1.13}
\end{align}
where $\omega(r)$ is the single polymer pair structure factor and $c(\vert\mathbf{r}-\mathbf{r}^{\prime}\vert)$ is the direct correlation function. The detailed derivation of the above
equation is given in Appendix \ref{appendixA}. We consider a small fluctuation in the density in the Fourier space and obtain
\begin{equation}
 \delta\hat{\rho}_p(\mathbf{k})\left[\frac{1}{n_p} - \hat{\omega}^2(\mathbf{k})\hat{c}(\mathbf{k})\right] + 
 \hat{\omega}^2(\mathbf{k})\delta\hat{\phi}(\mathbf{k}) = 0.
 \label{eq1.14}
\end{equation}
Comparing the above equation equation with the definition of the response function in equation \eqref{eq1.12} we get 
\begin{equation}
 \hat{\chi}_p(\mathbf{k}) = -\frac{\hat{\omega}^2(\mathbf{k})}{\frac{1}{n_p} - \hat{\omega}^2(\mathbf{k})\hat{c}(\mathbf{k})}.
 \label{eq1.15}
\end{equation}
Note that in the above derivation the three polymer correlations $\delta c(\vert\mathbf{r}-\mathbf{r}^{\prime}\vert)/
\delta\rho_p(\mathbf{r}^{\prime\prime})$ have been neglected. We can get rid of the direct correlation function in the 
response function in the above equation using the PRISM equation \eqref{eqA1.11}, which relates the direct correlation function
$c(r)$ to the pair correlation function $g(r)$. The Fourier transform of the equation is
\begin{align}
 \hat{h}(\mathbf{k}) = \omega^2(\mathbf{k})\hat{c}(\mathbf{k}) + n_p\omega(\mathbf{k})\hat{c}(\mathbf{k})\hat{h}(\mathbf{k}),
 \label{eq1.16}
\end{align}
 where $h(r) = g(r) - 1$. Using this equation in the static structure factor $S_p(\mathbf{k}) = \omega(\mathbf{k}) + 
 n_p\hat{h}(\mathbf{k})$, we see that the response function is related to the static structure factor by
\begin{equation}
 \hat{\chi}_p(\mathbf{k}) = - n_p\omega(\mathbf{k})S_p(\mathbf{k}). 
 \label{eq1.17}
\end{equation}
This is the static fluctuation-dissipation theorem for polymers in dimensionless form. The point particle version of the theorem \cite{hansen1990theory}
is recovered by setting $\omega(\mathbf{k}) = 1$. Thus for salt we have $\hat{\chi}_s(\mathbf{k}) = - n_sS_s(\mathbf{k})$. 
Thus for polymers we see that the fluctuations (structure factor) are the product 
of the single polymer fluctuations and the inter-polymer fluctuations. This form of the fluctuations was originally proposed
in a phenomenological way in Ref \cite{hayter1980correlations}.
Plugging them into the response function in equation \eqref{eq1.11}, the EP becomes
\begin{equation}
 \hat{\phi}(\mathbf{k}) = \frac{\Gamma\hat{\rho_f}(\mathbf{k})}{k^2 + k_s^2S_s(\mathbf{k}) + 3\Gamma\omega(\mathbf{k}) S(\mathbf{k}) },
 \label{eq1.18}
\end{equation}
where $k_s^2 = 4\pi\Gamma q_sn_s$. Similar expressions for EPs were obtained phenomenologically for Gaussian
polyelectrolytes by Khokhlov and Khachaturian \cite{khokhlov1982theory}, and later within random phase approximation
by Boryu and Erukhimovich \cite{boryu1988statistical}. Writing the EP in this form, which has been derived in the linear response
regime, allows us to go to the stronger coupling regimes (beyond linear response) easily using the machinery of the integral
equations. In Section \ref{Sec3} we explicitly work out the EP incorporating stronger correlations beyond the 
linear response regime.

\section{Mean field limit}
\label{Sec2}
\begin{figure*}[h]
        \centering
           \subfloat{%
              \includegraphics[height=6.2cm]{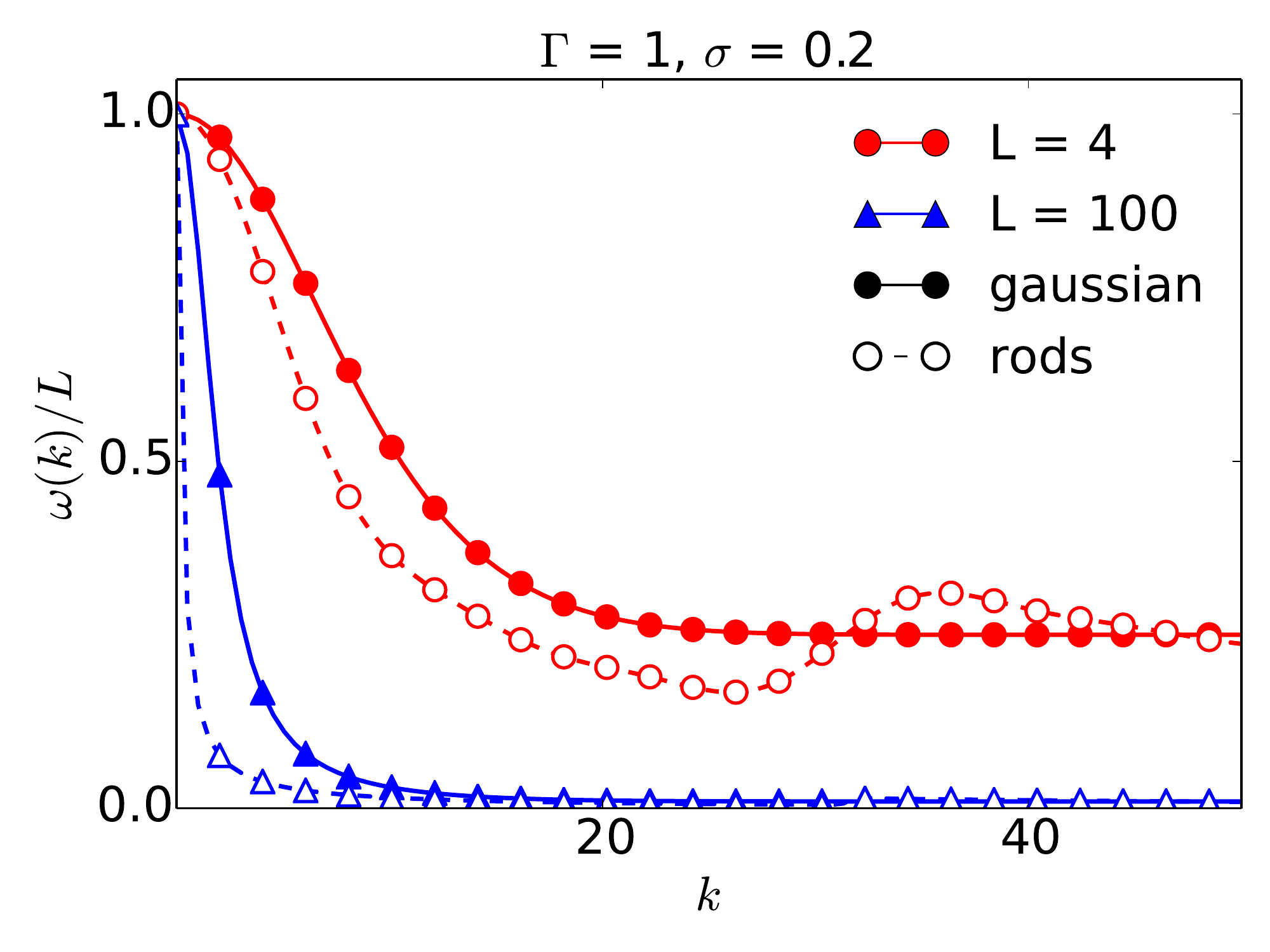}%
           }
           \subfloat{%
              \includegraphics[height=6.2cm]{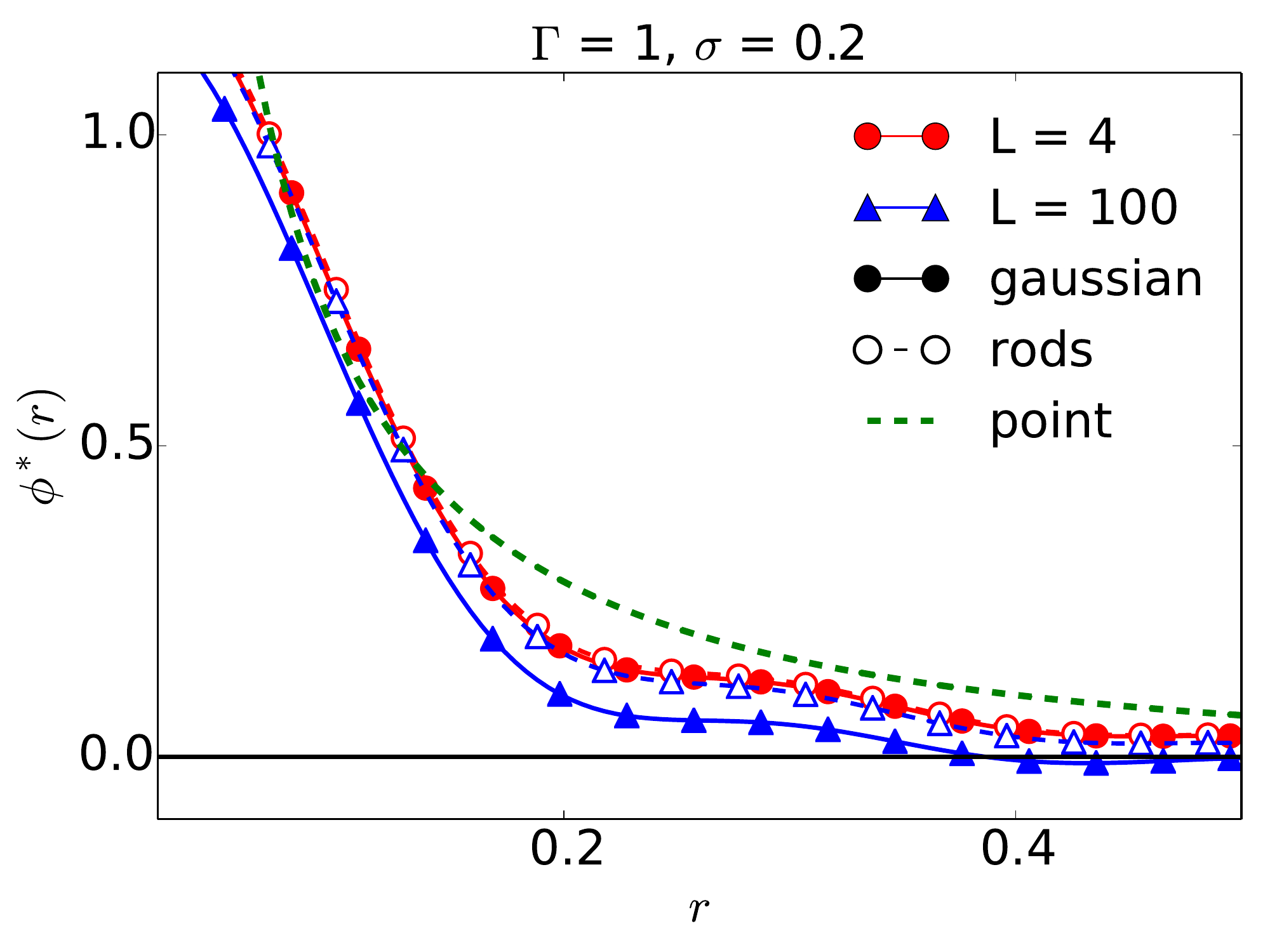}%
           }
           \caption{(a) The static structure factor for the GPEs (solid) and the RPEs (dashed) for polymer lengths $L = 4$ and $100$. 
           (b) EPs from a point charge for GPEs (solid), RPEs(dashed) and point (solid, no markers). For PEs, we consider two polymer lengths $L = 4$ and $100$.}
           \label{Fig1}
 \end{figure*}

In this Section we obtain the EP in the presence of a point charge $Q$ in the mean field approximation.
The fixed charge distribution function is then $\hat{\rho}_f(\mathbf{k}) = Q$. 
In the mean field limit $S_p(\mathbf{k}) = S_s(\mathbf{k}) = 1$, hence the EP scaled by $Q\Gamma$ is given by
\begin{align}
 \hat{\phi}^{\ast}(\mathbf{k}) = \frac{1}{Q\Gamma}\hat{\phi}(\mathbf{k})  = \frac{1}{k^2 + k_s^2 + 3\Gamma\omega(\mathbf{k}) }.
 \label{eq2.1}
\end{align}
In the point particle limit, obtained by taking $L \rightarrow 1$, we recover the Yukawa potential with an inverse
screening length $\sqrt{k_s^2 + 3\Gamma}$. 
The EP thus captures the screening effects modifying the long ranged Coulomb potential in case of point charges.
For GPEs the single polymer structure factor is given by \cite{schweizer1994prism}
\begin{align}
 \hat{\omega}(\mathbf{k}) = \left(1 - f(k)^2 - 2f(k)/L + 2f(k)^{L+1}/L\right)/(1-f(k))^2,
 \label{eq2.2}
\end{align}
where $f(k) = \exp(-k^2\sigma^2/6)$. For RPEs the corresponding structure factor is \cite{shew1997integral}
\begin{equation}
 \hat{\omega}(\mathbf{k}) = 1 + \frac{2}{L}\sum\limits_{j=1}^{L-1}(L-j)\frac{\sin jk\sigma}{jk\sigma}.
 \label{eq2.3}
\end{equation}
The plots of the single polymer structure factor scaled by the length of the polymers $L$ are shown in Figure \ref{Fig1}-(a)
for both the RPEs and GPEs. At large $\mathbf{k}$, $\omega(\mathbf{k})\rightarrow 1$ and $\omega(\mathbf{k} = 0) = L$. Longer polymers
have longer ranged correlations in position space and hence are short ranged in momentum space.

\begin{figure*}[h]
        \centering
           \subfloat{%
              \includegraphics[height=6.2cm]{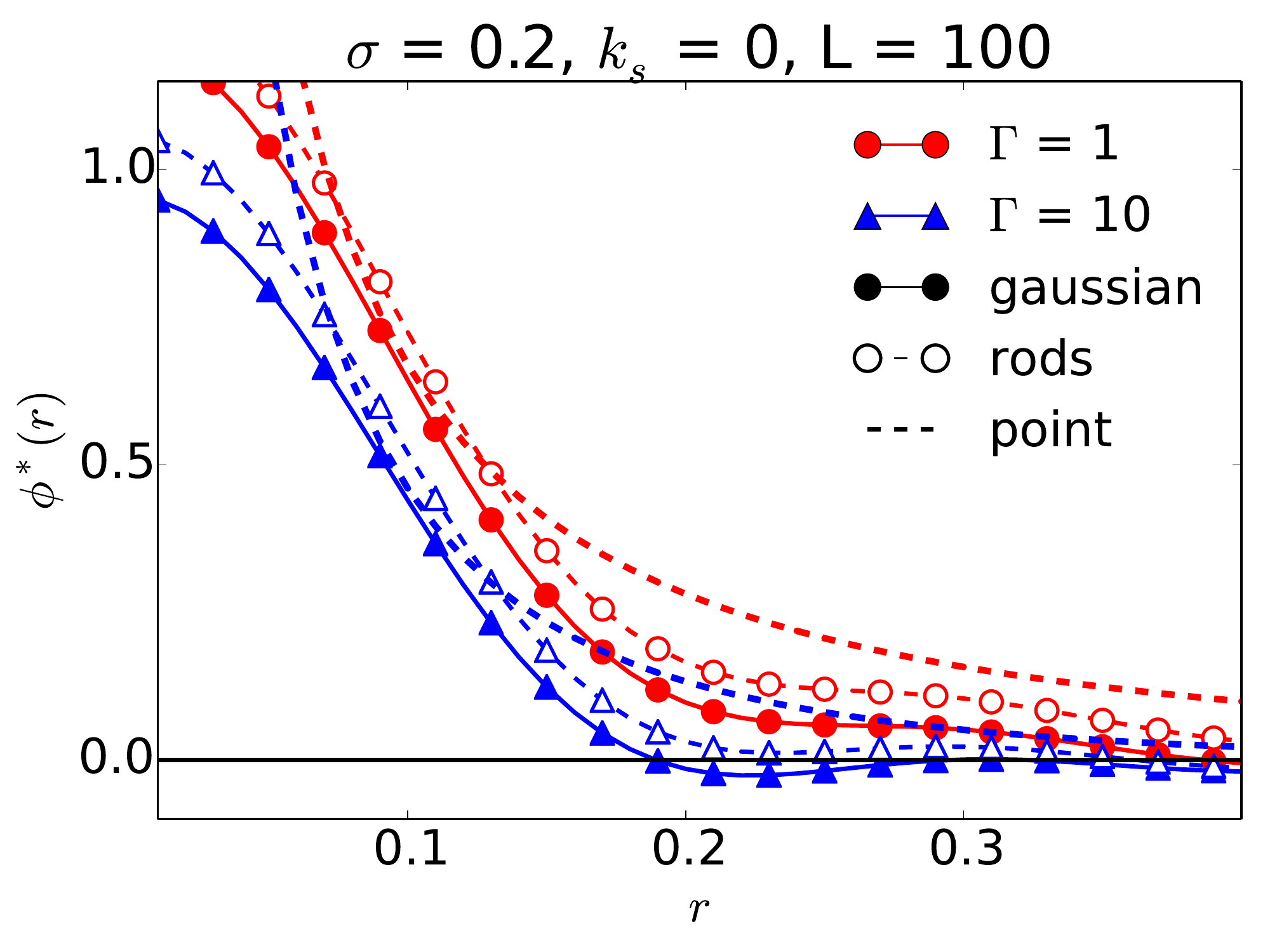}%
           }
           \subfloat{%
              \includegraphics[height=6.2cm]{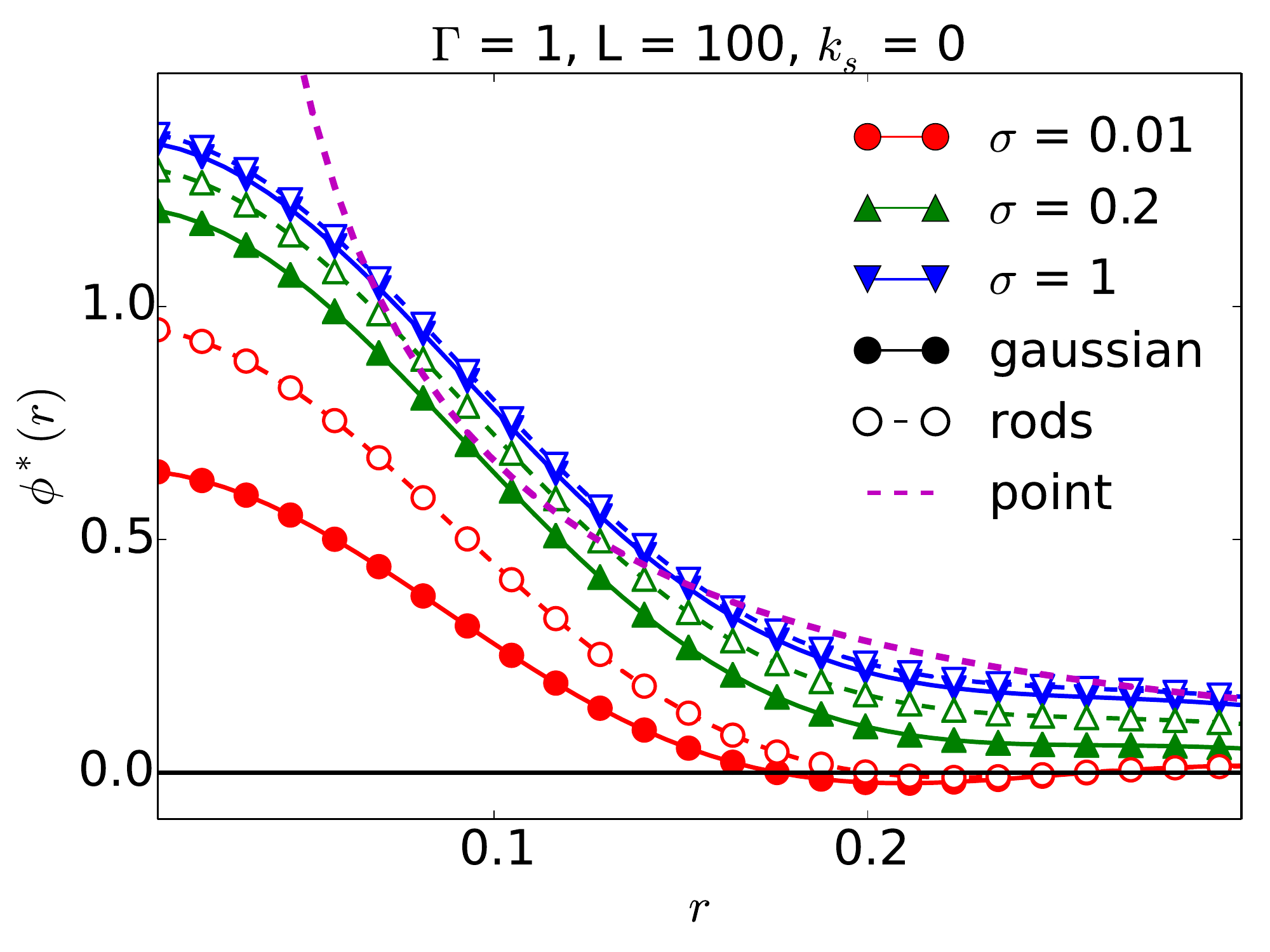}%
           }
           \caption{(a) EPs from a point charge for GPEs (solid), RPEs(dashed) and point (solid, no markers) for Coulomb couplings $\Gamma = 1$ and $10$. 
           (b) EPs for different monomer lengths $\sigma = 0.01$, $0.2$ and $1$ for GPEs and RPEs.}
           \label{Fig2}
 \end{figure*}
 
Figures \ref{Fig1}-(b) shows the scaled EP as defined in equation \eqref{eq2.1} for RPEs and GPEs of lengths $4$ and
$100$ respectively at $\Gamma = 1$.
At short distances the EPs for polymers are finite unlike the Yukawa potential of the point particles. In the other words
the Coulomb singularity at the origin is softened for polymers. The EPs have a weaker dependence 
on the polymer length especially for the RPEs as seen from Figure
\ref{Fig1}-(b). This behavior can be understood from the fact that the integrand of the EP vanishes at $\mathbf{k} = 0$ because of 
the 3D measure and at large $k$, $\hat{\phi}(\mathbf{k})\sim \frac{1}{k^2}$ because $\omega(k)\rightarrow 1$. Thus the EPs are almost similar for all polymers
except at intermediate distances.  While the EP for point charges is a Yukawa potential which is always repulsive, 
the EP can become attractive for longer polymers more so at higher $\Gamma$s. Unlike the point particles, even in the 
mean field limit the EPs can develop an attractive region 
due to intra-polymer correlations that are present in the EP as seen from equation \eqref{eq2.1}.
The EP reflects the deviation of the potential of a fixed charge distribution from the Coulomb potential due to the 
correlations. The correlations (fluctuations)
tend to lower the energy of the system. When the correlations are sufficiently high, the EPs can become attractive.
Attractions caused by fluctuations are in fact the principal mechanism of the like-charge attractions \cite{ha1997counterion,pincus1998charge}.
For GPEs the single polymer correlations are stronger than the RPEs. 
Because of the flexibility of the GPEs, the monomers can easily reorient 
themselves inside the chains and have more orientational degrees of freedom to lower their energies than the rigid RPEs.
Therefore the EP for GPEs becomes attractive at lower $\Gamma$s than the rods as seen in Figure \ref{Fig2}-(a). 
In the Figure, the tail region of the EPs show a weak oscillatory behavior at large $\Gamma$ especially for the GPEs as obtained by earlier simulations
\cite{turesson2007interactions} and theories \cite{boryu1988statistical}.
Figure \ref{Fig2}-(b) shows that the polymers with smaller monomer lengths $\sigma$ have lower EPs. They have a higher charge density and hence stronger
correlations which lower the EPs. For the flexible GPEs with smaller monomers the energy is lowered further than the RPEs.  

In the threadlike limit (polymer thickness goes to zero), we can deduce an analytic form for the EPs. In this limit the intra-molecular structure factor can be 
written in the form $\omega(\mathbf{k}) \approx 1/\left(L^{-1} + k^2\sigma^2/12\right)$. The EP in equation \eqref{eq2.1} 
in particular in the long polymer limit and low salt limit $k_s \approx 0$ reads
\begin{equation}
 \phi^{\ast}(r) = \frac{1}{4\pi r}\exp\left(-\frac{\sqrt{3}}{(\sigma^2/\Gamma)^{1/4}}r\right)\cos\left(\frac{\sqrt{3}}{(\sigma^2/\Gamma)^{1/4}}r\right).
\end{equation}
This form of the EPs was obtained by Mutukumar \cite{muthukumar1996double} using a field theoretic argument at low salt concentrations. 

\begin{figure*}[h]
        \centering
           \includegraphics[height=6.5cm]{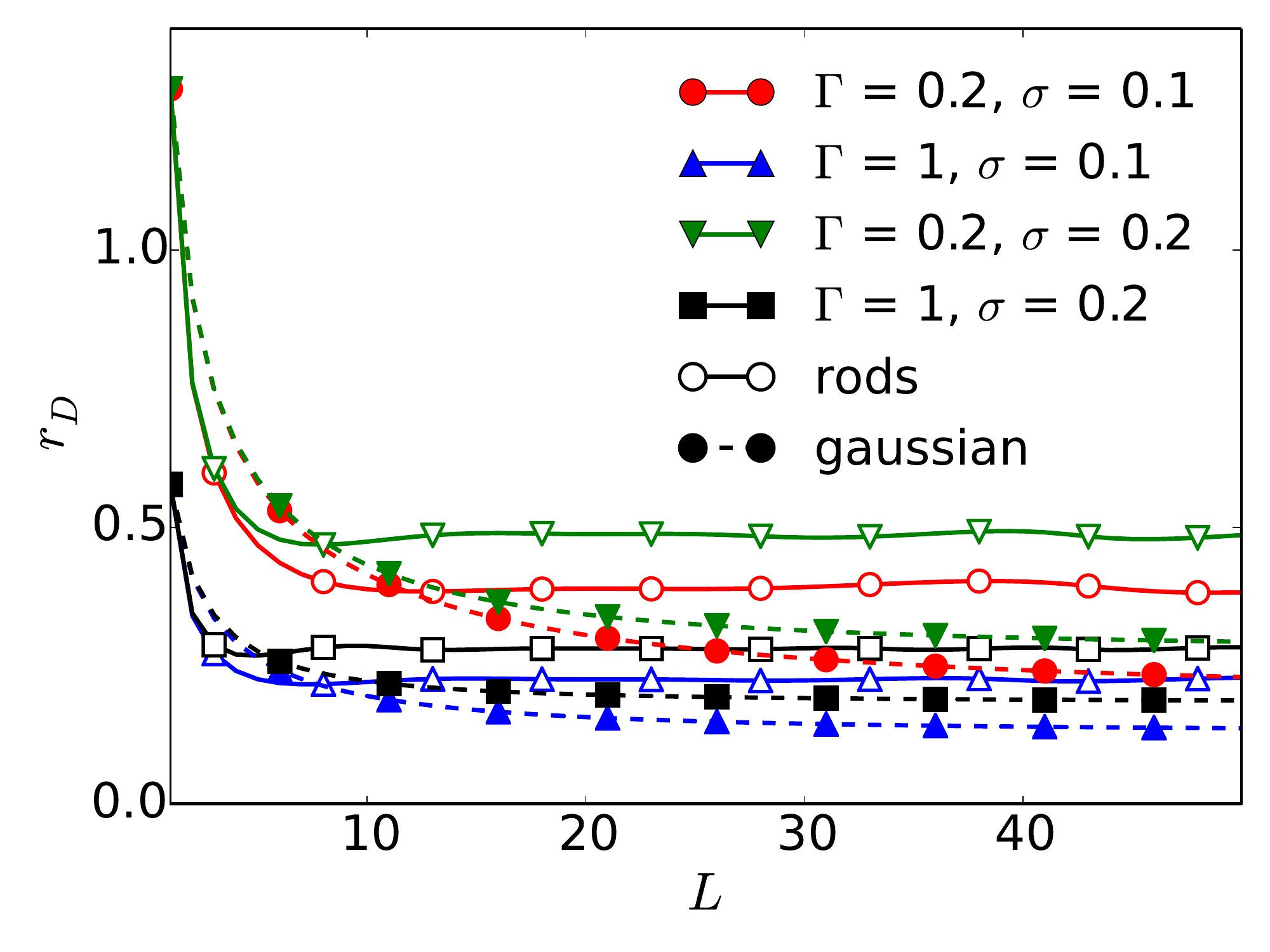}%
           \caption{The Debye length vs polymer length $L$ for Coulomb couplings $\Gamma = 0.2$ and $1$, and monomer lengths $\sigma = 0.1$ and $0.2$ for GPEs (dashed)
           and RPEs (solid). }
           \label{Fig3}
 \end{figure*}

The polymers also screen charges more effectively than point charges. This is seen from Figures \ref{Fig1}-(b) and \ref{Fig2}. We can 
get a quantitative estimate of the screening through the Debye length, $r_D$. The Debye length measures the distance to which the influence of 
a charge persists in the medium. Smaller Debye length implies that the charges are strongly screened in the medium. The Debye length can be obtained 
from equation \eqref{eq2.1} by solving the following equation self-consistently
\begin{equation}
 r_D = 1/\sqrt{3\Gamma\omega(1/r_D)}.
\end{equation}
For point charges this become $r_D = 1/\sqrt{3\Gamma}$. We plot the Debye length vs the length of the polymers in Figure \ref{Fig3}
for various monomer lengths and $\Gamma$s. Longer and thinner polymers have smaller Debye length and hence they screen the point charge more strongly
then the shorter polymers. However further increasing the polymer length does not change the Debye length because of the steric effects.
Also screening is stronger for GPEs than RPEs. Since $L = 1$ corresponds to point charge, we see from the Figure that screening is stronger for 
polymers than point ions.

\section{Beyond mean field}
\label{Sec3}

\begin{figure*}[h]
        \centering
           \subfloat{%
              \includegraphics[height=6.5cm]{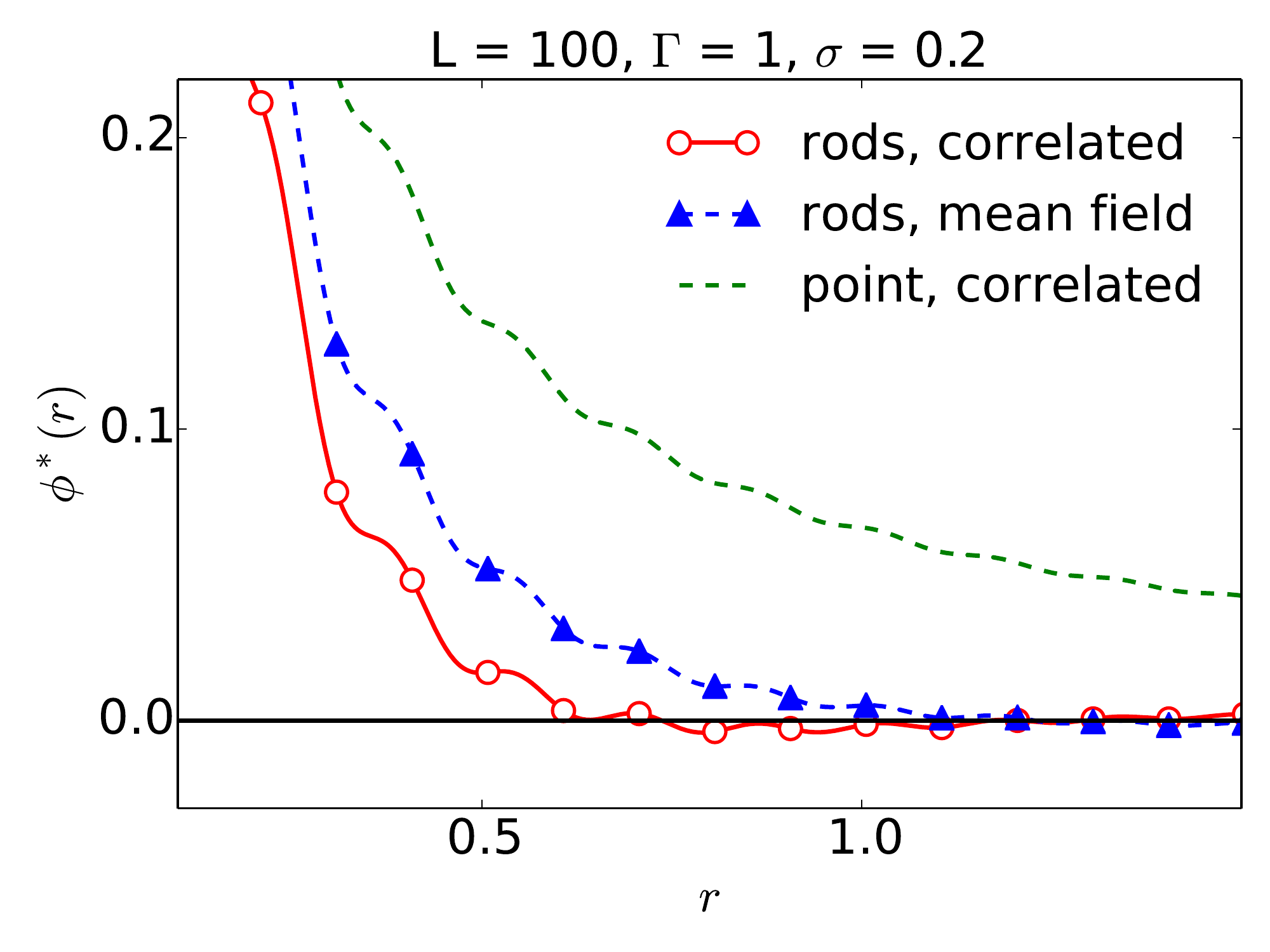}%
           }
           \caption{EP from a point charge for RPEs with the full correlations, calculated from equations \eqref{eq3.1} and \eqref{eq3.2}, at $\Gamma = 1$.}
           \label{Fig4}
 \end{figure*}

\begin{figure*}[h]
        \centering%
           \includegraphics[height=6.5cm]{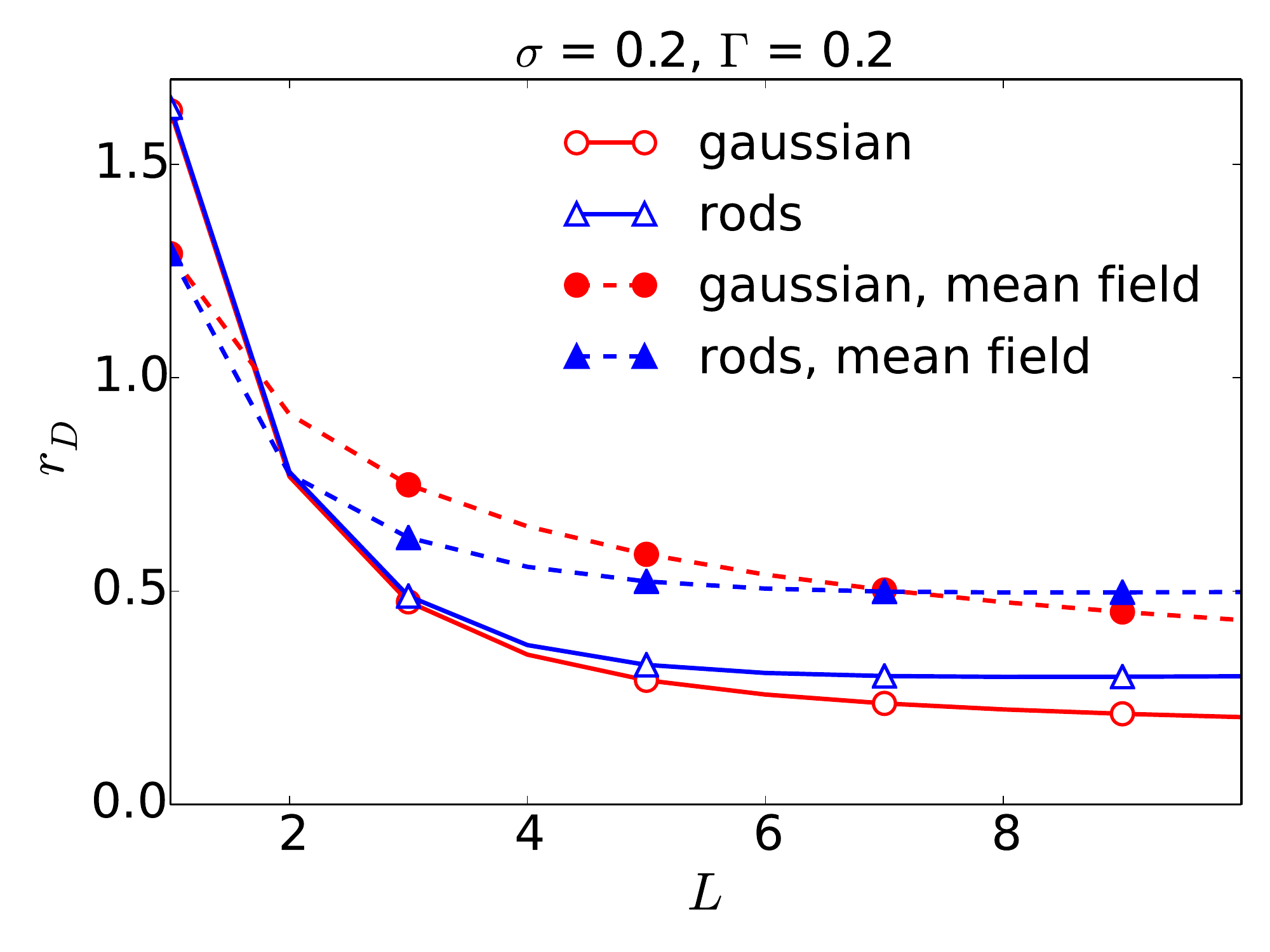}%
           \caption{The Debye length vs polymer length $L$ for GPEs (circle) and RPEs (triangle) showing both the mean field (dashed) and correlated
           results (solid).}
           \label{Fig5}
 \end{figure*}

To obtain the EP beyond the mean field, we calculate the direct correlation function self-consistently using the PRISM equation \cite{schweizer1994prism}
\begin{equation}
 g(r) - 1 = \int d\mathbf{r}^{\prime}d\mathbf{r}^{\prime\prime}\omega(\vert\mathbf{r}-\mathbf{r}^{\prime}\vert)c(\vert\mathbf{r}^{\prime}-\mathbf{r}^{\prime\prime}\vert)
 \omega(r^{\prime\prime}) + \int d\mathbf{r}^{\prime}d\mathbf{r}^{\prime\prime}\omega(\vert\mathbf{r}-\mathbf{r}^{\prime}\vert)c(\vert\mathbf{r}^{\prime}-\mathbf{r}^{\prime\prime}\vert)
 \bar{\rho}h(r^{\prime\prime}),
 \label{eq3.1}
\end{equation}
and the LWC formalism \cite{laria1991reference}
\begin{align}
 \ln g(r) = -\int d\mathbf{r}^{\prime}d\mathbf{r}^{\prime\prime}\omega(\vert\mathbf{r}-\mathbf{r}^{\prime}\vert)\beta V(\vert\mathbf{r}^{\prime}-\mathbf{r}^{\prime\prime}
 \vert)\omega(r^{\prime\prime}) + h(r) -\int d\mathbf{r}^{\prime}d\mathbf{r}^{\prime\prime}\omega(\vert\mathbf{r}-\mathbf{r}^{\prime}\vert)c(\vert\mathbf{r}^{\prime}-\mathbf{r}^{\prime\prime}\vert)
 \omega(r^{\prime\prime}), 
 \label{eq3.2}
\end{align}
following the procedure given in Reference \cite{shew1997integral}.
In Figure \ref{Fig4} we plot the EP from equation \eqref{eq1.18} with the full correlations.
Including the intra-polymer correlations further lowers the EP than the corresponding mean field EP as seen in the Figure.
Because of the finite size the polymer-polymer correlations are stronger
for polymers than that of point charges. Figure \ref{Fig5} shows the Debye length for both the GPEs and RPEs in the mean field and after the polymer-polymer
correlations are included. The point particle limit ( $L = 1$ ) and weak coupling, the structure factor $S(k)\approx k^2/(k^2+1/r_D^2)$
and from this the Debye length becoms $r_D \approx \sqrt{2/3\Gamma}$ whereas in the mean field $r_D = 1/\sqrt{3\Gamma_D}$. Therefore
the Debye length is lower in the mean field for point particle in Figure \ref{Fig5}. But as the length of the polymers increases the 
screening length for the corelated case decreases faster than the mean field case, implying that inter-polymer correlations cause
stronger screening. 

\section{Conclusion and discussions}
\label{Sec4}
Using the linear response theory in the Poisson-Boltzmann equation, we have derived an EP of a fixed charge distribution in a polyelectrolytic
solution. We have calculated the response function of the polymers from the integral equation for the density of polymers. With the help of the PRISM
equation we relate the response function to the static structure factor of the polymers. This relationship is the static fluctuation-dissipation
theorem for polymers, where the fluctuations (structure factor) splits into the fluctuations within the polymers and inter-polymer fluctuations.
The EP can be calculated once the structure factors are known. 

In the mean field limit, the inter-polymer structure factor is 1. In this limit we have calculated the EP over a range of lengths of the 
polymers, the Coulomb couplings and the salt concentrations. In the low salt and high salt regime our EP coincides with the EP obtained by 
Mutukumar using field theoretic arguments. Close to the origin the EP for polyelectrolytes is repulsive but is finite unlike the case
for point charges where it is singular. On increasing the polymer length or the Coulomb coupling, the EP gradually develops an attractive
regime at short distances still having a Debye-Huckel repulsive form at longer distances. At very large polymer lengths the EP has an
oscillatory behavior. The attractive part of the potential is a result of the configurational entropy and the connectivity of the polymers.
The GPEs have more configurational entropy which results in their EP becoming more attractive then the RPEs. The polyelectrolytes screen 
charges better than the point charges. We use the Debye length to get the quantitative estimate of the screening by the polyelectrolytes.
On increasing the polymer length the screening rapidly increases at short polymer length but does not change at larger lengths. This is 
because of steric effects most part of the large polymers can not penetrate within certain distance of the test charge and so no 
further screening is possible after a certain polymer length. 

Using the PRISM and LWC equations we have obtained the inter-polymer correlations to calculate the EP beyond the mean field. 
Including these correlations further increases the attractive region of the EP compared to the mean field case. In fact the EP becomes
attractive at Coulomb coupling as low as $\Gamma \approx 0.2$. The Debye length also decreases on including the correlations, implying
stronger screening.

The GPEs and RPEs are the two extreme limits for the semi-flexible polymers. Most realistic biological systems are made of semi-
flexible polymers. To accurately characterize the experiments in biopolymers we need to extend our analysis to the semi-flexible 
polymers. This would add a new parameter, the rigidity of the polymers, to the problem. But then we can smoothly move between 
the two limits: GPEs when the rigidity is zero to RPEs when the rigidity is infinite. This would be done in a subsequent paper.

\section{Acknowledgement}

S.D. thanks P. Benetatos for his valuable comments and discussions.

\appendix

\section{Integral equation for polymer density}
\label{appendixA}
In this Section we phenomenologically derive an expression for the monomer density in terms of an external potential using
the RISM model developed by Chandler \textit{et al} \cite{chandler1986density2}.
The RISM model relates the polymer site density $\alpha$, $\rho_{\alpha}(\mathbf{r})$ to the intra-molecular pair correlation function 
$\omega_{\alpha\beta}(\vert\mathbf{r}-\mathbf{r}^{\prime}\vert)$, the local chemical
potential $\psi_{\alpha}(\mathbf{r}) = \mu_{\alpha} - \phi_{\alpha}(\mathbf{r})$ and the direct correlation function 
$c_{\alpha\beta}(\vert\mathbf{r}-\mathbf{r}^{\prime}\vert)$ 
(note we use the direct correlation function of uniform system for simplicity)
\begin{align}
\rho_{\alpha}(\mathbf{r}) = \prod\limits_{\gamma\neq\alpha}\omega_{\alpha\gamma}*\exp(f_{\gamma}),
 \label{eqA1.3}
\end{align}
where 
\begin{equation}
f_{\gamma} = \psi_{\gamma} + \sum_{\eta}c_{\gamma\eta}*\rho_{\eta}.
\label{eqA1.35}
\end{equation}
The symbol $\ast$ denotes the convolution operation $p * q = \int d\mathbf{r}^{\prime}p(\mathbf{r})q(\vert\mathbf{r}-\mathbf{r}
^{\prime}\vert)$ and have dropped the position dependence to keep notations simple. Like the PRISM \cite{schweizer1994prism} formalism
we replace the quantities at each site by the corresponding site averaged quantity. Summing over the index $\alpha$ and replacing
$\omega_{\alpha\gamma}$ by $\omega = \frac{1}{L}\sum_{\alpha,\gamma=1}^L\omega_{\alpha\gamma}$ we get
\begin{align}
 \rho_p  & = \sum_{\alpha}\rho_{\alpha} \approx \prod\limits_{\gamma}\omega*\exp(f_{\gamma})
 \label{eqA1.4}
\end{align}
For polyatomic systems RHS of equation \eqref{eqA1.3} should have an additional convolution with the site-averaged pair correlations $\omega$
\cite{chandler1986density2}
\begin{align}
 \ln\rho_p & \approx \sum_{\gamma}\ln\left(\omega*\exp(f_{\gamma})*\omega\right).
 \label{eqA1.5}
\end{align}
Expanding the exponential on RHS of the above equation and keeping till the first order term we get
\begin{align}
 \ln\rho_p & \approx \sum_{\gamma}\ln\left( 1 + \omega*f_{\gamma}*\omega\right) \nonumber\\
 & \approx  \omega*\sum_{\gamma}f_{\gamma}*\omega \nonumber \\
 & = \omega*f*\omega.
 \label{eqA1.6}
\end{align}
In the first step of the derivation we have made use of the identity $\int d\mathbf{r}\omega(\mathbf{r}) = 1$.
Using the explicit form of $f$ in equation \eqref{eqA1.35} the final expression of the equilibrium density becomes
\begin{equation}
 \ln\rho_p = \omega*\psi*\omega + \omega*c*\rho_p*\omega,
 \label{eqA1.7}
\end{equation}
where $\psi = \sum_{\alpha}\psi_{\alpha}$ and $\rho_p = \sum_{\alpha}\rho_{\alpha}$. We use 
Percus's idea to obtain an expression for the pair correlation function \cite{hansen1990theory}.
When one of the polymers is fixed at the origin, it would act as an external potential. In this case $\psi(r) = V(r)$
and the density becomes the pair correlations $\rho_p(r) = n_pg(r)$ \cite{hansen1990theory}. Plugging these in 
equation \eqref{eqA1.7} we get
\begin{equation}
 \ln g = \omega*(-\beta V)*\omega + n_p\omega*c*(g-1)*\omega.
 \label{eqA1.8}
\end{equation}
Using the PRISM equation \cite{schweizer1994prism} 
\begin{equation}
g - 1 = \omega * c * \omega + n_p\omega * c * (g - 1),
 \label{eqA1.85}
\end{equation}
we see that equation \eqref{eqA1.8} is identical to the HNC formalism of Laria, Wu, and Chandler (LWC) \cite{laria1991reference} for molecular systems,
except for an extra convolution of $\omega$ in the second term on the RHS. To make our theory consistent with the LWC formalism we drop the last term
on the RHS of equation \eqref{eqA1.7}, the convolution with $\omega$. Now we put the distance dependence in equation \eqref{eqA1.7} explicitly
 \begin{align}
  \ln\left(\rho_p(\mathbf{r})\lambda^3/z\right) = -\int d\mathbf{r}^{\prime}d\mathbf{r}^{\prime\prime}\omega(\vert\mathbf{r}-\mathbf{r}^{\prime}\vert)\beta\phi(\vert\mathbf{r}^{\prime}-\mathbf{r}^{\prime\prime}\vert)
 \omega(r^{\prime\prime}) + \int d\mathbf{r}^{\prime}d\mathbf{r}^{\prime\prime}\omega(\vert\mathbf{r}-\mathbf{r}^{\prime}\vert)
 c(\vert\mathbf{r}^{\prime}-\mathbf{r}^{\prime\prime}\vert)\rho_p(\mathbf{r}^{\prime\prime}),
 \label{eqA1.9}
 \end{align}
where $\lambda = \sqrt{h^2/2\pi mk_BT}$ is the thermal wavelength and $z$ is the fugacity of the system.
 
The direct correlation function in equation \eqref{eqA1.9} is self-consistently from the LWC equation 
\begin{align}
 \ln g(r) = -\int d\mathbf{r}^{\prime}d\mathbf{r}^{\prime\prime}\omega(\vert\mathbf{r}-\mathbf{r}^{\prime}\vert)\beta V(\vert\mathbf{r}^{\prime}-\mathbf{r}^{\prime\prime}
 \vert)\omega(r^{\prime\prime}) + h(r) -\int d\mathbf{r}^{\prime}d\mathbf{r}^{\prime\prime}\omega(\vert\mathbf{r}-\mathbf{r}^{\prime}\vert)c(\vert\mathbf{r}^{\prime}-\mathbf{r}^{\prime\prime}\vert)
 \omega(r^{\prime\prime}), 
 \label{eqA1.10}
\end{align}
and the PRISM equation 
\begin{equation}
 g(r) - 1 = \int d\mathbf{r}^{\prime}d\mathbf{r}^{\prime\prime}\omega(\vert\mathbf{r}-\mathbf{r}^{\prime}\vert)c(\vert\mathbf{r}^{\prime}-\mathbf{r}^{\prime\prime}\vert)
 \omega(r^{\prime\prime}) + \int d\mathbf{r}^{\prime}d\mathbf{r}^{\prime\prime}\omega(\vert\mathbf{r}-\mathbf{r}^{\prime}\vert)c(\vert\mathbf{r}^{\prime}-\mathbf{r}^{\prime\prime}\vert)
 n_ph(r^{\prime\prime}),
 \label{eqA1.11}
\end{equation}
where $n_p = \frac{1}{V}\int d\mathbf{r}\rho_p(\mathbf{r})$ and $h(r) = g(r) -1$.

\bibliography{refs}
%\begin{thebibliography}{99} 
%\end{thebibliography}

\bigskip

\end{document}